\documentclass[showpacs,superscriptaddress,floatfix,12pt,amsmath,tightenlines]{revtex4}

\usepackage{graphicx}
\usepackage{psfrag}

\newcounter{saveeqn}

\begin{document}


\title{Reaction coordinates for the flipping of genetic switches}

\author{Marco J. Morelli}
\email{morelli@amolf.nl}
\affiliation{FOM Institute for Atomic and Molecular Physics, Kruislaan 407, 1098 SJ Amsterdam, The Netherlands}

\author{Sorin T\u{a}nase-Nicola}
\affiliation{FOM Institute for Atomic and Molecular Physics, Kruislaan 407, 1098 SJ Amsterdam, The Netherlands}

\author{Rosalind J. Allen}
\affiliation{SUPA, School of Physics, University of Edinburgh, James Clerk Maxwell Building, The King's Buildings, Mayfield Road, Edinburgh EH9 3JZ, UK}

\author{Pieter Rein ten Wolde}
\affiliation{FOM Institute for Atomic and Molecular Physics, Kruislaan 407, 1098 SJ Amsterdam, The Netherlands}

\date{\today}

\begin{abstract}
 
  We present a detailed analysis, based on the Forward Flux Sampling
  (FFS) simulation method, of the switching dynamics and stability of
  two models of genetic toggle switches, consisting of two
  mutually-repressing genes encoding transcription factors (TFs); in
  one model (the exclusive switch), the two transcription factors
  mutually exclude each other's binding, while in the other model
  (general switch) the two transcription factors can bind
  simultaneously to the shared operator region.  We assess the role of two
  pairs of reactions that influence the stability of these switches:
  TF-TF homodimerisation and TF-DNA association/dissociation.  In both
  cases, the switch flipping rate increases with the rate of TF
  dimerisation, while it decreases with the rate of TF-operator
  binding. We factorise the flipping rate $k$ into the product of the
  probability $\rho(q^*)$ of finding the system at the dividing
  surface (separatrix) between the two stable states, and a kinetic
  prefactor $R$.  In the case of the exclusive switch, the rate of
  TF-operator binding affects both $\rho(q^*)$ and $R$, while the rate
  of TF dimerisation affects only $R$. The general switch displays a
  higher flipping rate than the exclusive switch, and both TF-operator
  binding and TF dimerisation affect $k$, $R$ and $\rho(q^*)$. To
  elucidate this, we analyse the transition state ensemble (TSE). For
  the exclusive switch, the TSE is strongly affected by the rate of
  TF-operator binding, but unaffected by varying the rate of TF-TF
  binding. Thus, varying the rate of TF-operator binding can
  drastically change the pathway of switching, while changing the rate
  of dimerisation changes the switching rate without altering the
  mechanism. The switching pathways of the general switch are highly
  robust to changes in the rate constants of both TF-operator and
  TF-TF binding, even though these rate constants do affect the
  flipping rate; this feature is unique for non-equilibrium systems.


\end{abstract}

\maketitle

\section{Introduction}\label{intro}
Biochemical networks with multiple stable states are omnipresent in
living cells. Multistability can provide cellular memory, it can
enhance the sharpness of the response to intra- and extracellular
signals, it can make the cell robust against biochemical noise, and it
allows cells to differentiate into distinct cell types. Although a
multistable biochemical network can flip between alternative states
due to random fluctuations (``noise''), in many cases the states are
very stable and the network typically only switches from one state to
the next under the influence of an external signal
\cite{gensigbook}. A key question, therefore, in understanding
multistable biochemical networks is what controls the stability of the
steady states. To answer this question, we have to elucidate the
pathways of switching between steady states. Switching events are,
however, intrinsically difficult to study experimentally, because the
switching event itself can be much faster than the typical life time
of the steady state. Computer simulations are a valuable tool for
studying biochemical networks, especially for rare processes such as
switching. However, precisely because such events are rare, special
techniques are required to simulate them. One such technique is
Forward Flux Sampling (FFS) \cite{Allen05,Allen06_1,Allen06_2}, and in
this paper, we use FFS in combination with committor distributions to
analyse in detail the effect of two important sources of
fluctuations---transcription factor dimerisation and transcription
factor-DNA binding---on the flipping rate and switching pathways of two
models of bistable genetic toggle switches. We hope that this analysis
may serve as a paradigm for studying multistable biochemical networks
as well as other rare events in non-equilibrium systems.

If a biochemical network is bistable, with two stable states $A$ and
$B$, respectively, then it will show a bimodal steady-state
probability distribution, $\rho(q)$, of some order parameter $q$. This
order parameter can be the concentration of a species, or a
combination of the concentrations of a number of species. It is
usually interpreted as a reaction coordinate that measures the
progress of the `reaction' from state $A$ to $B$. Recently, such
bimodal distributions have been measured experimentally for
biochemical networks~\cite{Gardner00,Ozbudak04,Acar05}. These
distributions are potentially useful, because they are linked to the
rate of switching from one state to the other. In particular, we have
recently shown \cite{Warren05} that not only for equilibrium systems,
but also for systems that are out of equilibrium such as biochemical
networks, the rate of switching from state $A$ to state $B$, $k_{AB}$,
can be written as the product of two factors:
\begin{equation}
\label{eq:kAB}
k_{AB} = R \rho  (q^*).
\end{equation}
Here, $q^*$ denotes the location of the dividing surface, the
separatrix \cite{Warren05,Walczak05}, which separates the two states
$A$ and $B$. The above relation is useful because it shows that the
rate of switching from one steady state to the next, is given by
the probability $\rho(q^*)$ of being at the dividing surface times a
kinetic prefactor $R$ that describes the average flux of trajectories
crossing the dividing surface. 
However, while the rate constant $k_{AB}$ does not depend on the
choice of the order parameter $q$ as long as it connects states $A$
and $B$, both $\rho(q^*)$ and $R$ do depend on the choice of $q$. If
$q$ is the ``true'' reaction co-ordinate that accurately describes the
switching process, then $q^*$ corresponds to the transition state and
$\rho(q^*)$ and $R$ provide accurate measures for the probability of
being at the transition state and the flux of trajectories leaving the
transition state for state $B$ \cite{Bolhuis00}. A key issue in
analysing rare events in general is therefore identifying the reaction
coordinate $q$ that accurately describes the progress of the
transition.

FFS can be used to compute $k_{AB}$, $\rho(q^*)$, $R$, and to generate
members of the transition path ensemble
\cite{Allen05,Allen06_1,Allen06_2,Valeriani07}. To identify the
reaction coordinate, the transition paths can be analysed using
committor distributions; this approach is commonly applied in the
field of soft-condensed matter physics \cite{Bolhuis00}, and we have
recently demonstrated how it can be used to analyse the transition
pathways of biochemical switches \cite{Allen05}. Each configuration
$x$ of our system has a commitment probability or ``committor'',
$P_B(x)$. This is the probability that a trajectory, fired at random
from that configuration, will reach state $B$ before state $A$. Given
$P_B(x)$, we can define the ``transition state ensemble'' (TSE)
\cite{DBG}, which is the collection of configurations along the
reaction pathways which have committor value $P_B(x) = 0.5$. We can
extract TSE configurations from our switching pathways by computing
committor values for all the points along the pathways and selecting
those points (several per path) with $P_B = 0.5$. We then try to find
an order parameter (or combination of order parameters) that
accurately describes these TSE configurations. To test likely order
parameters, one can compute the probability distribution for the order
parameter for the TSE configurations \cite{Bolhuis00,Allen05}. Poorly
chosen order parameters will show a broad or even bimodal distribution
\cite{ Geissler99}, while good order parameters will show a narrow
distribution of values in the TSE.

In this paper, we apply FFS to study two different models of genetic
toggle switches, consisting of two genes $A$ and $B$ that mutually
repress each other
\cite{Kepler01,Aurell02,Zhu04,Warren04,Warren05,Allen05,Lipshtat06,Ushikubo06,Loinger07}.
The genes encode transcription factors (TF) ${\rm{A}}$ and ${\rm{B}}$
respectively. These can form homodimers, in which form they can bind
to a regulatory region of the DNA (represented by an operator site
$O$) and regulate transcription. The dimer ${\rm{A_2}}$ represses the
transcription of $B$ when bound to $O$ and vice versa (see
Fig. \ref{fig:diagram}).  In the first model, called the {\em
  exclusive switch} \cite{Warren04, Warren05} the dimers of the two
species are not allowed to simultaneously bind to the operator; in the
second model, called the {\em general switch}, the operator
can bind both types of homodimers at the same time \cite{Warren04,Warren05}.
Both switches have one stable state in which
${\rm{A}}$ is abundant, and ${\rm{B}}$ scarce, and another stable
state in which ${\rm{B}}$ is abundant and ${\rm{A}}$ scarce.
We simulate the switch using the Gillespie algorithm
~\cite{Gillespie77}, in combination with FFS. The Gillespie Algorithm
 is a widely used and efficient
Kinetic Monte Carlo scheme~\cite{Bortz75} for chemical reactions,
which generates trajectories in correspondence with the chemical
master equation. 

Switching events are driven by random fluctuations. Key fluctuation
sources in this network are TF-TF and TF-DNA association and
dissociation reactions. By varying the rates of these reactions, while
keeping their equilibrium constants fixed, we can vary independently
the time scales and hence the correlation times of these
fluctuations. The correlation times of fluctuations are important, since
they determine the extent to which the fluctuations propagate in the
network \cite{Paulsson05,Shibata05,TanaseNicola06}.

We therefore begin by calculating how the stability of both switches
varies with the rate of TF-TF association and dissociation
(dimerisation), and with the rate of TF-operator association and
dissociation (operator binding).  We vary the association and dissociation
rates together, keeping their ratio ({\em i.e.} the equilibrium
constant) unchanged.  The switching rate is strongly affected: for
both models, $k_{AB}$ decreases as the rate of operator binding
increases, and increases as the rate of dimerisation increases.
Analysing the effects on $\rho(q^*)$ and $R$, we find that the two
models behave differently: for the exclusive switch, the rate of
operator binding changes both $\rho(q^*)$ and $R$, while the rate of
dimerisation affects only $R$; for the general switch, the
rate of dimerisation affects both $\rho(q^*)$ and $R$, while the rate
of operator binding predominantly changes $\rho(q^*)$.

We then show that the effect of TF-TF and TF-DNA fluctuations on $k$,
$R$, and $\rho(q^*)$ can be understood by elucidating the switching
mechanism using commmittor distributions. We find that for the
exclusive switch the difference in total copy number number of the two
species is not a complete reaction coordinate: the state of the
operator is also an important factor in determining the committor
value \cite{Allen05}. In contrast, we find little evidence that
dimerisation is an important ingredient of the reaction coordinate.
This explains why the rate of operator binding affects both the
probability of being at the separatrix and the kinetic prefactor,
while dimerisation only affects the kinetic prefactor. For the general
switch, the situation is markedly different: the switching mechanism
is highly robust to changes in both the rate of operator binding and
the rate of dimerisation. Hence, changing these rate constants does
not change the route the switching pathways take in state space, yet
does affect the flipping rate. This is a manifestation of the fact
that this is a non-equilibrium system---in an equilibrium system the
switching rate cannot be changed without changing the switching
pathways. The general implication of this observation is that in order
to understand the stability of biochemical switches, we need to
understand not only the composition of the transition state ensemble,
but also the dynamics of the transition paths.

In the next Section, we describe the model genetic switches in more
detail. In the subsequent Section, we briefly discuss the FFS
technique. We then present the results on the switching rate, the
kinetic prefactor and the probability of being at the separatrix for
both models, showing how they depend on the rates of operator binding
and of dimerisation.  In the next Sections, we discuss switching
pathways and reaction coordinates first for the exclusive switch, and
then for the general switch.  We end with a discussion of the
implications of our findings for the modelling of multistable
biochemical networks and the study of rare events in other
non-equilibrium systems.

\section{Models: the Exclusive and the General Switch}
\label{sec:GS}

We consider a genetic toggle switch consisting of two genes, each of
which represses the other
\cite{CA,Kepler01,Warren04,Warren05,Allen05,Lipshtat06}. A switch of
this kind has been constructed and shown to be bistable in {\em
  E. coli} \cite{Gardner00}. We study both the general
switch and the exclusive switch, introduced by one of us
\cite{Warren04, Warren05}. The general switch is represented by the
following set of reactions \cite{Warren04,Warren05,Allen05}:
\begin{subequations}
\label{eq:original_set}
\begin{align}
\mathrm{A}+\mathrm{A} \underset{k_{\rm b}}{\overset{k_{\rm f}}\rightleftharpoons} \mathrm{A}_2, &\quad &\mathrm{B}+\mathrm{B} \underset{k_{\rm b}}{\overset{k_{\rm f}}\rightleftharpoons} \mathrm{B}_2\\
{O} + \mathrm{A}_2 \underset{k_{\rm off}}{\overset{k_{\rm on}}\rightleftharpoons} {O}\mathrm{A}_2, &
\quad & {O} + \mathrm{B}_2 \underset{k_{\rm off}}{\overset{k_{\rm on}}\rightleftharpoons} {O}\mathrm{B}_2 \\
{O}\mathrm{A}_2 + \mathrm{B}_2 \underset{k_{\rm off}}{\overset{k_{\rm on}}\rightleftharpoons} {O}\mathrm{A}_2\mathrm{B}_2, &
\quad & {O}\mathrm{B}_2 + \mathrm{A}_2 \underset{k_{\rm off}}{\overset{k_{\rm on}}\rightleftharpoons} {O}\mathrm{A}_2\mathrm{B}_2 \\
{O} {\overset{k_{\rm prod}}\to} {O} + \mathrm{A}, & \quad &  {O} {\overset{k_{\rm prod}}\to} {O} + \mathrm{B}\\
{O}\mathrm{A}_2 {\overset{k_{\rm prod}}\to} {O}\mathrm{A}_2 + \mathrm{A}, &
\quad &
{O}\mathrm{B}_2 {\overset{k_{\rm prod}}\to} {O}\mathrm{B}_2 + \mathrm{B}  \\
\mathrm{A} {\overset{\mu}\to} \emptyset, & \quad &\mathrm{B} {\overset{\mu}\to} \emptyset
\end{align}
\end{subequations}
In this reaction scheme, $O$ represents a DNA regulatory sequence
adjacent to two divergently transcribed genes $A$ and $B$.  These code
respectively for proteins A and B, as shown in
Fig. \ref{fig:diagram}. Genes $A$ and $B$ can each randomly produce
proteins with the same rate, but whether they do so depends on the
state of the operator $O$. Proteins A and B can each form a homodimer
that can bind to the operator. When an ${\rm{A_2}}$ dimer is bound to
$O$, the production of B is blocked, and likewise, when a ${\rm{B_2}}$
dimer is bound to $O$, the production of A is blocked. When both
dimers are bound to the operator, no protein can be produced.
Proteins can also vanish (in the monomer form), modelling degradation
and dilution in a cell.  This model can be modified by removing
reactions (\ref{eq:original_set}c): in this case, transcription
factors mutually exclude each other's binding to the operator. We
refer to the switch described by the whole set of chemical reactions
(\ref{eq:original_set}) as the ``general switch''; the ``exclusive
switch'' consists of the same set of reactions, except for reactions 
(\ref{eq:original_set}c). 

We have assumed in this model 
that transcription, translation and protein folding can be modelled
as single Poisson processes, neglecting the many 
substeps that lead to the production of a
protein. Ref. \cite{Warren05} discusses the effects of both shot noise
and fluctuations in the number of proteins produced per mRNA
transcript on the switch stability. We also note here that
while mean-field analysis predicts that cooperative binding of the TFs
to the DNA is essential for bistability \cite{CA}, it has recently been
demonstrated that bistability can be achieved without cooperative binding when the discrete nature of the reactants is taken into
account \cite{Lipshtat06}.

We choose $k_{\rm prod}^{-1}$ as the unit of time for our simulations,
and we use the volume of the system, $V$, as the unit of volume. In
this paper, we will use the following ``baseline'' set of parameters:
$k_{\rm f}=5k_{\rm prod}V$, $k_{\rm b}=5k_{\rm prod}$ (so that $K_{\rm
  D}^d=k_{\rm b}/k_{\rm f}=1/V$), $k_{\rm on}=5k_{\rm prod}V$, $k_{\rm
  off}=k_{\rm prod}$ (so that $K_{\rm D}^b=k_{\rm off}/k_{\rm
  on}=1/(5V)$), $\mu=0.3k_{\rm prod}$. These parameters are chosen to
be representative of typical cellular values, as discussed in Section
\ref{sec:dis}. For simplicity, the model switches are completely
symmetrical - rate constants for equivalent reactions involving A and
B are the same. The mean field analysis performed in \cite{Warren05}
demonstrates analytically for both systems the existence of three
fixed points for the parameter values listed above: two symmetrical
stable states, one rich in A and the other rich in B, separated by one
unstable state where the total number of A equals the total number of
B. This implies that the system can be considered a truly bistable
switch. However, while this analysis indicates the regions in
parameter space where the system is bistable, it cannot predict the
switch stability nor elucidate the switching pathways. For this
reason, we carry out stochastic Kinetic Monte Carlo simulations using
the Gillespie algorithm \cite{Gillespie77,Bortz75}. In previous work,
we have shown that the switch stability depends strongly on the mean
copy number of species A and B \cite{Warren04}, which is given by the
ratio of the protein production and decay rates, $k_{\rm
  prod}/\mu$. In this paper, we investigate its dependence on the
other parameters $k_{\rm f}, k_{\rm b}, k_{\rm on}$ and $k_{\rm off}$,
which govern key sources of fluctuations in the network - TF-TF and
TF-DNA association and dissociation reactions.

\section{Method: Forward Flux Sampling} \label{sec:FFS}
Conventional simulation methods are ineffective for studying rare events such as the flipping of biochemical switches, because the vast majority of the computational effort is spent in simulating the uninteresting waiting
times in between the events. For this reason, specialised methods are
required, and we have recently developed the Forward Flux Sampling
(FFS) technique \cite{Allen05,Allen06_1,Allen06_2}. FFS is well suited
to simulating biochemical networks, since unlike most other rare event
methods \cite{BCDG}, it can be used for out-of-equilibrium systems. In this paper, we use FFS to calculate rate constants, transition paths and stationary probability distribution functions for the model genetic switch.

To obtain the rate constant $k_{AB}$ for a rare transition between two states $A$ and $B$, FFS exploits the fact
that (in steady state) $k_{AB}$ can be written as
the product of two factors:
\begin{equation}
\label{k_FFS}
k_{AB} = \Phi_A P_{AB}
\end{equation}
Here, $\Phi_A$ is the number of trajectories that leave state $A$ per
unit time, while $P_{AB}$ is the conditional probability
that these trajectories subsequently reach state $B$ without returning to $A$. An order parameter $\lambda$ must be chosen, which defines states $A$ and $B$: if $\lambda < \lambda_0$ the system is in state $A$,
while it is in state $B$ if $\lambda > \lambda_n$. The parameter $\lambda$ is then used to further define a series of nonintersecting
interfaces $\{\lambda_1,\dots,\lambda_{n-1}\}$, with $\lambda_i <
\lambda_{i+1}$, such that any trajectory from $A$ to $B$ has to cross all the interfaces $\{\lambda_0,\dots,\lambda_{n}\}$, without reaching $\lambda_{i+1}$ before it has crossed $\lambda_i$. The conditional probability $P_{AB}$ can be written as 
\begin{equation}
\label{eq:P_AB}
P_{AB} = \prod_{i=0}^{n-1} P(\lambda_{i+1}|\lambda_i), 
\end{equation}
where $P(\lambda_{i+1}|\lambda_i)$ is the conditional probability that a trajectory
that comes from $A$ and has crossed $\lambda_i$ for the first time, will
subsequently reach $\lambda_{i+1}$ before returning to $A$. Several different algorithms can be used to calculate $\Phi_A$, $P(\lambda_{i+1}|\lambda_i)$ and to obtain transition paths; in this paper, we have used the original scheme \cite{Allen05,Allen06_1}. Briefly, one first performs a
conventional (``brute-force'') simulation to compute $\Phi_A$, which is the
number of times that the trajectory crosses $\lambda_0$, coming from $\lambda <
\lambda_0$, divided by the total simulation time. When one of these crossings occurs, the configuration of the system is stored, so that this simulation run generates not only an estimate for $\Phi_A$,
but also a collection of points at interface $\lambda_0$. In
the next stage, one chooses a point at random from this collection, and fires off a new
trajectory from this point, which is continued until the system either reaches the
next interface $\lambda_1$ or returns to state $A$. If $\lambda_1$ is reached, the system configuration at $\lambda_1$ is stored in a new collection. This procedure is repeated a number
of times until a sufficiently large number of points at the next
interface has been generated. An
estimate for $P(\lambda_1|\lambda_0)$ is obtained from the number of trials which reached $\lambda_1$, divided by the total number of trials fired from $\lambda_0$. Starting from the new collection of points at $\lambda_1$, one then repeats this whole procedure to drive the system to $\lambda_2$, and so on. Eventually, the system reaches state $B$, upon which the rate constant can be calculated from Eqs.(\ref{k_FFS}) and (\ref{eq:P_AB}). Furthermore, a (correctly weighted) collection of trajectories corresponding to the transition (``transition paths'') can be obtained by tracing back those trial paths that arrive in $B$ to their origin in $A$. 

We have recently shown \cite{Valeriani07} that this procedure can be used to generate not only the rate constant and transition paths, but
also the stationary distribution $\rho(q)$, as a function of a chosen order parameter (or parameters) $q$. This is achieved by continuously
updating a histogram in the parameter $q$ during the trial run procedure, as described in Ref. \cite{Valeriani07}. To obtain  $\rho(q)$, histograms for the ``forward'' ($A \to B$) and ``backward''  ($B \to A$) transition must be combined. However, since our model switch is symmetric, the two histograms are identical in this case. The parameter $q$ does not have to be the same as $\lambda$, although in this paper we have chosen $q=\lambda$.

In FFS, a series of interfaces are used to drive the system over a
``barrier'', in a ratchet-like manner. The efficiency of the method of
course depends on the choice of the order parameter $\lambda$, the
positioning of the interfaces, number of trials etc
\cite{Allen06_2}. However, it is important to note that $\lambda$ does
not have to be the true reaction coordinate for the transition. The
choice of $\lambda$ does not impose any bias on the system dynamics:
transition paths are free to follow any possible path between state
$A$ and $B$. The choice of $\lambda$ should not affect the computed
rate constant, transition paths or $\rho(q)$. Furthermore, the FFS
method does not make a Markovian assumption about the transition
paths, or any assumptions about the distribution of state points at
the interfaces $\{\lambda_0,\cdots,\lambda_n\}$: each point at
interface $i$ lies on a true dynamical path which originates in the
initial state $A$. This turns out to be essential for the model
genetic switch.

For the FFS calculations presented in this paper, we have chosen as
$\lambda$ parameter the difference between the total copy numbers of
the two proteins: $\lambda \equiv n_A - n_B$, with $n_X \equiv N_{\rm
  X}+2N_{\rm X_2}+2N_{\rm OX_2}$ the total copy number of species X =
A or B in the exclusive switch and $n_X \equiv N_{\rm X}+2N_{\rm
  X_2}+2N_{\rm OX_2} + 2N_{\rm OA_2 B_2}$ the total copy number of
species X = A or B in the general switch.

\section{Results}
Key sources of fluctuations in in this reaction network are TF-TF and
TF-DNA association and dissociation reactions. We can vary the
influence of these fluctuations on the network dynamics by changing
the rate constants for association and dissociation, keeping the
equilibrium constant (the ratio of association and dissociation rate
constants) fixed, so that the macroscopic dynamics of the system
remain unchanged. When these reactions are fast, fluctuations are
short-lived on the timescale of the slower protein production and
degradation reactions, so that the effect of a fluctuation is lost
over just a few production/degradation reactions. However, for slow
association-dissociation reactions, fluctuations in, for example, the
ratio of monomers to dimers, can persist over the timescale of a
series of production/decay reactions, and thus have a strong influence
on the dynamics of the whole network. In what follows, we first
discuss the effects of varying the rates of operator binding and
dimerisation on the switching rate for both genetic switch models, and then, 
to elucidate these effects, we separately discuss the switching pathways for the two cases.

\subsection{Switching rates} \label{sec:sr}
Figs. \ref{fig:sw_rates_ES}A and \ref{fig:sw_rates_ES}B show the
flipping rate $k_{AB}$ for the exclusive switch as a function of the
dimerisation rate $k_{\rm f}$ and the operator binding rate $k_{\rm
  on}$, respectively (keeping the dissociation constants
constant). The results for the general switch are shown in
Figs. \ref{fig:sw_rates_GS}A and \ref{fig:sw_rates_GS}B.  It is clear
that for both switches the two sources of fluctuation have very
different effects on the stability: while $k_{AB}$ increases with the
rate of dimerisation (panels A), it decreases with the rate of
operator binding (panels B). Thus, fluctuations in the TF-DNA
association/dissociation reactions tend to destabilise the switch,
whereas (counter-intuitively), fluctuations in the TF-TF
association/dissociation reactions {\em increase} the switch
stability.

To understand the origin of these effects, we factorise the flipping rate $k_{AB}$ into the product of the probability of finding the system
at the dividing surface $\rho(q^*)$ and a kinetic prefactor $R$, as in Eq. (\ref{eq:kAB}). Fig. \ref{fig:Eq} shows the steady-state
probability distribution $\rho(\lambda)$ of finding the system at a
particular value of the order parameter $\lambda$, for different values
of the dimerisation and operator binding rate: panel A refers to the exclusive switch and panel B to the general switch.
These distributions were computed using FFS, as described in Section \ref{sec:FFS} and Ref. \cite{Valeriani07}. 

We first note that both distributions exhibit peaks at
$\lambda\approx\pm 27$, corresponding to the two stable
states. Secondly, the locations of the two stable states and the shape
of the stationary distributions are fairly insensitive to both the
rate of dimerisation and the rate of operator binding. However, around
$\lambda=0$ the distributions, especially that of the general switch,
are much more sensitive to changes in the rate
constants. Interestingly, the probability of finding the system at the
value $\lambda=0$ is markedly differently for
the two models: while for the exclusive switch $\rho(\lambda)$
exhibits a minimum, representing an unstable steady state for the
system, in the case of the general switch, the probability
distribution has a local maximum, indicating the presence of a
metastable steady state \cite{Warren05,Lipshtat06}.
Finally, we note that for an equilibrium system, fluctuations do not
influence the stationary distribution function: the effect of $k_{\rm
  f}$ and $k_{\rm{on}}$ on $\rho(q)$ is a clear characteristic of the
non-equilibrium nature of the dynamics in this system.  

From the distribution $\rho(q)$, we compute the probability of being
at the minimum of the curve, $\rho(q^*)$. For the exclusive switch,
this point corresponds to the dividing surface $\rho(\lambda = 0)$; we
show in Figs. \ref{fig:sw_rates_ES}C and \ref{fig:sw_rates_ES}D how
this quantity varies with $k_{\rm f}$ and $k_{\rm on}$,
respectively. In the case of the general switch, the transition
happens through the metastable state at $\lambda=0$. However, the
rate-limiting step for the flipping is to get to the minimum of
$\rho(\lambda)$, which is now located at $\lambda\approx\pm
5$. Therefore, for the general switch, $\rho(q^*)$ was computed for
$q^* = \lambda = 5$; it is shown in Figs. \ref{fig:sw_rates_GS}C and
\ref{fig:sw_rates_GS}D.  By combining $\rho(q^*)$ with Eq. (\ref{eq:kAB}), we
compute the kinetic prefactor $R$, shown in panels E and F of
Figs. \ref{fig:sw_rates_ES} and \ref{fig:sw_rates_GS}, for the
exclusive and general switch, respectively.

We observe that for the exclusive switch $\rho(q^*)$ depends upon the
operator binding rate (Fig. \ref{fig:sw_rates_ES}D), but not upon the
rate of dimerisation (Fig. \ref{fig:sw_rates_ES}C), while
for the general switch $\rho(q^*)$ depends upon both rate
constants (Figs. \ref{fig:sw_rates_GS}C,D). In both models, the
kinetic prefactor $R$ increases with the
rate of dimerisation (Figs. \ref{fig:sw_rates_ES}E,\ref{fig:sw_rates_GS}E), while it decreases with the rate of operator
binding (albeit much less so in the general switch; Figs. \ref{fig:sw_rates_ES}F,\ref{fig:sw_rates_GS}F).  One might expect
that a change in $\rho(q^*)$ reflects a change in the location of the
switching pathways in state space. This would suggest that in the
general switch, the switching pathways depend upon both the rate of
dimerisation and the rate of operator binding, while for the exclusive
switch the switching mechanism does depend upon the rate of operator
binding, but not on the rate of dimerisation. While the conclusion for
the exclusive switch is correct, that for the general
switch is not, as we discuss in the next two sections.

\subsection{Switching pathways - Exclusive switch}
To understand the effects of the operator binding and dimerisation
fluctuations on the switching rate, we would like to determine what
the true reaction coordinate is for the switching process and whether
it involves these fluctuations. To do this, we need to examine the
transition paths for switching, which are also generated by FFS. We
will focus on three sets of parameters: (1) the base-line set, with
operator binding rate $k_{\rm on} = 5k_{\rm prod}V$ and dimerisation
rate $k_{\rm f} = 5 k_{\rm prod}V$; (2) a set with slow dimerisation,
$k_{\rm f} = 0.1k_{\rm prod}V$, and $k_{\rm on} = 5k_{\rm prod}V$; (3)
a set with fast operator binding, $k_{\rm on} = 500k_{\rm prod}V$,
$k_{\rm f} = 5k_{\rm prod}V$. As above, in all cases the dissociation
rates are scaled such that the equilibrium constants remain constant:
$K_{\rm D}^d=k_{\rm b}/k_{\rm f}=1/V$ and $K_{\rm D}^b=k_{\rm
  off}/k_{\rm on}=1/(5V)$. In this section, we discuss the exclusive
switch, while the next section focusses on the general switch.

To analyse the progress of the system as it flips from one state to
the other, we have averaged the switching trajectories in the $P_B$
ensemble. The committor $P_B(x)$ is the probability that a trajectory
propagated at random from configuration $x$ reaches state $B$ before
state $A$. The $P_B$ {\em ensemble} is formed by those configurations $x$
that have the same value of $P_B$; $\langle Q (x) \rangle_{P_B}$ thus
denotes the average of a quantity $Q(x)$ in the ensemble of
configurations $x$ with the same value of $P_B$. Given an ensemble of
switching paths obtained with FFS, we can harvest configurations $x$
with the same value of $P_B$. Indeed, each transition path has at
least one configuration for every value of $P_B$.  $P_B(x)$ can be
used to characterise the progress of the transition from $A$ to
$B$---in a sense, it is the true reaction coordinate and our task is
to identify coordinates that characterise $P_B$.  However, its
evaluation is computationally very expensive.  We have computed $P_B$
for configurations in the transition paths that were generated using
FFS, by firing 100 test trajectories from each configuration.  The
average paths in the $P_B$ ensemble are rather ``noisy'', precisely
because $P_B$ is a stochastic quantity that has to be estimated by a
computationally demanding procedure.

Fig. \ref{fig:NANB}A shows the average switching pathways for the
exclusive switch in the $n_{\rm A},n_{\rm B}$ plane, where $n_{\rm A}$
and $n_{\rm B}$ are the total copy numbers of species A and B,
respectively ($n_{\rm A}=N_{\rm A}+2N_{\rm A_2}+2N_{\rm OA_2}$ for the
exclusive switch and $n_{\rm A}=N_{\rm A}+2N_{\rm A_2}+2N_{\rm OA_2}
+2N_{\rm OA_2B_2}$ for the general switch; similarly for B).
Paths are shown for both the $A \to B$ (solid lines) and $B \to A$ (dashed lines) transitions.   
Considering the red and black lines in Fig. \ref{fig:NANB}A, we see
that the dimerisation rate $k_{\rm{f}}$ has little effect on the
switching pathways (at least in this representation), while on
considering the green and black lines, it is clear that the operator
binding rate $k_{\rm{on}}$ does strongly influence the switching
pathways, especially in the region of the dividing surface, where
$n_{\rm A}=n_{\rm B}$: the pathways shift to lower values of $n_{\rm
  A}$ and $n_{\rm B}$ when $k_{\rm{on}}$ is increased.

Since it appears from Fig. \ref{fig:NANB}A that the state of the
operator is likely to play an important role in the switching
mechanism, we plot in Fig. \ref{fig:OB2_lambda}A the probability that
the operator is bound by a ${\rm B}_2$ dimer, $\langle N_{\rm OB_2}
\rangle$ as a function of $\lambda$. 
Comparing the solid black and red lines, we see that
changing the rate of dimerisation has indeed little effect on the transition
paths. In contrast,
a comparison of the black and green solid lines shows that changing
the rate of operator binding has a strong effect on the switching
pathways. This indicates that operator state fluctuations
play an important role in switch flipping \cite{Allen05,Walczak05} -
so that the reaction coordinate depends not only on the difference in
the number of protein molecules, $\lambda$, but also on which protein
is bound to the operator. 

This fact is further illustrated in Fig. \ref{fig:histo}, which shows
histograms for configurations in the TSE of the transition from $A$ to
$B$; members of the TSE are points along the transition paths for
which $P_B=0.5$. Each panel in Fig. \ref{fig:histo} corresponds to a
different parameter set---the baseline parameter set in in the centre,
slow dimerisation on the left and fast operator binding on the
right. In each case, we divide the collection of TSE configurations
according to the state of the operator. For each operator state, we
plot histograms for the $\lambda$-coordinate, in such a way that the
area under a histogram corresponds to the total number of TSE points
with that operator state. The histograms are colour coded according to
operator state. Considering first the central panel of the upper row
(baseline parameter set), we see that the green histogram (${
  O{\rm{A_2}}}$) is shifted towards larger values of $\lambda$ than
the red histogram (${O{\rm{B_2}}}$)--- {\em{i.e.}} the state of the
operator and $\lambda$ are {\em correlated} in the TSE. This means
that if a $B$ dimer is bound to the operator, then, on average, the
number of $A$ molecules has to exceed the number of $B$ molecules in
order to have the same value of $P_B$, and vice versa. We also see
that the area under the ${O{\rm{B_2}}}$ histogram is larger than that
under the ${ O{\rm{A_2}}}$ histogram---indicating that the TSE has
predominantly ${\rm{B_2}}$ bound to the operator, even though the
switch is symmetric. Turning next to the right panel---rapid operator
association and dissociation---we see that again the $O{\rm{A_2}}$
histogram is shifted towards larger values of $\lambda$ relative to
the ${O{\rm{B_2}}}$. However, in this case, the areas under the two
histograms are approximately equal. Thus increasing the rate of
operator binding appears to have caused the transition state for
switch flipping to become symmetric in A and B. The left panel shows
the results for slow dimerisation, $k_{\rm{f}}=0.1$. This plot is
virtually indistinguishable from the baseline parameter
results---indicating that changing the dimerisation rate has little
effect on the transition state ensemble, as suggested by Fig.
\ref{fig:OB2_lambda}A. These results unambiguously demonstrate that,
for the exclusive switch, fluctuations arising from TF-DNA
association-dissociation reactions are central to the flipping
mechanism, while those arising from TF-TF association-dissociation
reactions have little effect on the flipping mechanism, although they
can influence the {\em dynamics} of the flipping trajectories and
hence the switching rate.

Drawing together the observations of Figs.
\ref{fig:sw_rates_ES}, \ref{fig:Eq}A, \ref{fig:NANB}A, \ref{fig:OB2_lambda}A and \ref{fig:histo}, we can now understand the
dependence of the exclusive switch flipping rate on the rate of operator binding
(Fig. \ref{fig:sw_rates_ES}B). In the limit of slow operator binding and
unbinding \cite{Allen05,Walczak05}, the binding of the minority
species to the operator strongly enhances the flipping of the switch:
when the minority species happens to bind the operator, it will stay
on the DNA for a relatively long time, thus blocking the synthesis of
the majority species and allowing the production of the minority
species. In this limit, the system can reach the dividing surface with
relatively few production/degradation events.  As the rate of operator
binding and unbinding is increased, each transition involves many
operator binding/unbinding events, and consequently proteins of both
species are produced and decay during the transition process. Here,
the state of the operator is increasingly slaved to the difference in
the total number of A and B molecules, $\lambda$. In the adiabatic
limit of fast operator binding, the probability that a molecule of
type $A$ or $B$ is bound to the operator is completely determined by
$\lambda$ \cite{Walczak05}. In this limit, the dividing surface is
located at $\lambda \approx 0$ and $\langle N_{\rm OA_2} \rangle
\approx \langle N_{\rm OB_2} \rangle$; to reach the separatrix, the
system has to wait for a series of fluctuations in the birth and decay
of both species that lead to $n_{\rm A} \approx n_{\rm B}$. This
implies that the total number of copies of A and B at the dividing
surface decreases as the rate of operator binding increases
(Fig. \ref{fig:NANB}A).  Because a series of production/decay events
are required to reach the separatrix, the probability $\rho(q^*)$ is
decreased as the rate of operator binding
increases (Fig. \ref{fig:sw_rates_ES}D). In addition, having reached the
separatrix, the system requires more production/decay events to take
it to the B state. This increases the probability that it will
``recross'' the separatrix and eventually return to $A$ instead of
contributing to $B$---resulting in a decrease in the kinetic prefactor
$R$ in Fig. \ref{fig:sw_rates_ES}F.

Figs. \ref{fig:NANB}--\ref{fig:histo} suggest that the rate of
dimerisation only has a marginal effect on the switching
pathways. However, our view of the switching pathways naturally
depends on the representation in which we choose to plot them. We have
investigated many representations to see whether the rate of
dimerisation could affect the switching pathways. Perhaps the most
important one is the average number of dimers $\langle N_{\rm B_2}
\rangle$ as a function of $\langle N_{\rm B}(N_{\rm B}-1)
\rangle$. However, also in this representation the rate of
dimerisation only has a very minor effect on the switching pathways;
in fact, near the top of the dividing surface, the dimerisation
reaction is in steady state (data not shown). This supports our
conlusion that dimerisation affects the rate at which the transition
paths traverse state space (and hence $R$), but not the route they
take (and thus not $\rho(q^*)$).

The effect of TF-TF association/dissociation fluctuations on the
dynamics of the trajectories can perhaps be understood by considering
that in order to start a switching event from one stable state to the
other, two copies of the minority species must be produced. They must
then dimerise and bind to the operator, to shut down production of the
majority species. If the dimerisation rate is comparable to the
degradation rate, it becomes increasingly probable that copies of the
minority species, once produced, are removed from the system before
they can form a dimer. Thus, decreasing the dimerisation rate actually
reduces the chance that the switch can flip. This effect is truly
dynamical in origin. We note that it is also fundamentally different
from enhanced switch stability via cooperativity due to nonlinear
degradation \cite{Buchler05}.

Lastly, while operator binding is an equilibrium reaction, it couples
to reactions that are out of equilibrium, such as protein production
and decay. As a result, the dynamics of operator binding can lead the
exclusive switch to behavior that is unique for non-equilibrium
systems. This can be seen by comparing the forward paths from $A$ to
$B$ with the backward paths from $B$ to $A$ in
Fig. \ref{fig:OB2_lambda}A. When the rate of operator binding is fast,
the forward and backward paths essentially coincide. This situation
differs markedly for the system with the base-line parameter set and
for the system with slow dimerisation: although the switch is
symmetric on interchanging $A$ and $B$, the transition path ensemble
(TPE) for the transition from $A$ to $B$ does not coincide with that
from $B$ to $A$ \cite{Allen05}. This is a manifestation of the fact
that this switch is a non-equilibrium system: for equilibrium systems
that obey detailed balance and microscopic reversibility, the forward
and backward paths must necessarily coincide.  The fact that the
forward and backward paths do not coincide also means that the
switching paths do not follow the path of highest steady-state phase
space density, which, for equilibrium systems, would correspond to the
lowest free-energy path: Since this system is symmetric, this
``lowest-free energy path'' is symmetric on interchanging species A
and B, while Fig. \ref{fig:OB2_lambda}A shows that the dynamical
switching trajectories are not (unless operator binding is fast). This
also means that for this system it is essential not to make the
Markovian assumption of memory loss, which underlies path sampling
schemes such as Milestoning \cite{Faradjian04} and PPTIS
\cite{Moroni04}.

\subsection{Switching pathways - General switch} \label{sec:sp_GS} We
now turn our attention to the switching pathways for the general
switch, again obtained with FFS. 
Fig. \ref{fig:NANB}B shows for the three different parameter sets the
switching trajectories as averaged in the $P_B$ ensemble and
projected onto the $n_{\rm A},n_{\rm B}$ plane.
As for the exclusive switch, the forward and backward paths do not
coincide, which, as mentioned above, reflects the fact that the
genetic switch is a non-equilibrium system. However, in many other
respects the pathways of the general switch differ markedly from those
of the exclusive switch. Firstly, the switching trajectories of the
general switch cross the dividing surface $\lambda=0$ at very low
values of $n_{\rm A}$ and $n_{\rm B}$---on average, there is only one
dimer of each species at the transition surface. Moreover, the paths
display a sharp deviation when they reach the dividing
surface. Lastly, paths obtained for different values of the rate
constants essentially coincide (the black, red and green lines
overlap). This last observation suggests that the transition paths are
rather insensitive to variations in the rate constants of dimerisation
and operator binding, an observation that should be contrasted with
the observation that both $\rho(q^*)$ and $R$ do depend upon the
magnitude of those rate constants (see Fig. \ref{fig:sw_rates_GS}).

Fig. \ref{fig:OB2_lambda}B shows the state of the operator for every
value of $\lambda$ during the transition, for the baseline parameter
set (the curves for the other parameter sets are virtually
indistinghuisable). Initially, when the system is still in the basin
of attraction of the stable state $A$, the operator is mostly in state
$O{\rm A_2}$. However, as the system leaves this basin, the state of
the operator rapidly becomes dominated by $O{\rm A_2B_2}$. Indeed,
this operator state, which is absent in the exclusive switch, plays a
pivotal role in flipping the general switch. Its occupancy peaks at
$\lambda \approx -5$, corresponding to the top of the barrier that
separates the stable state $A$ at $\lambda=-27$ from the metastable
state at $\lambda=0$. Here, at $\lambda=0$, the occupation statistics
of the operator is given by the equilibrium distribution $[O{\rm
  A_2}]$:$[O{\rm B_2}]$:$[O{\rm A_2B_2}]$, with $[O{\rm A_2}]=[O{\rm
  B_2}]$. As Fig. \ref{fig:histo}B shows, the transition state
ensemble coincides with the metastable state around $\lambda=0$, and
in this ensemble the operator is predominantly in state $[O{\rm
  A_2B_2}]$. As the system leaves the metastable state towards the $B$
state, the state of the operator progressively moves toward $[O{\rm
  B_2}]$.

We are now able to explain the process of flipping the general switch.
When a dimer of the minority species is produced, it immediately binds
to the operator and drives it in the state $O{\rm A_2B_2}$. In this
state, the production of {\em both} proteins is suppressed, and the
system is depleted of almost all its components
\cite{Warren04,Lipshtat06}.  The approach to the transition state is
then driven mostly by a decrease of the majority species via protein
decay rather than an increase of the minority species via protein
production, as in the exclusive switch; this is the reason why the
general switch crosses the diving surface at lower values of $n_A$ and
$n_B$ than the exclusive switch. Importantly, if one of the two dimers
leaves the operator, it can immediately rebind, thereby restoring the
previous situation and allowing the transition to continue.  By
contrast, if the minority species leaves the operator in the exclusive
switch, then most likely the majority species will bind the operator,
thereby blocking further progress of the transition. This explains why
both the pathways and the rate of flipping are much more
sensitive to the rate of operator binding in the exclusive switch than
in the general switch.

The presence of the state $O{\rm A_2B_2}$ also underlies the
metastability of the general switch at $\lambda=0$
(Fig. \ref{fig:Eq}B). As long as both species are present in the
system, the state $O{\rm A_2B_2}$ is the most stable operator state,
and in this state no proteins can be produced. As a consequence, a
small fluctuation in $\lambda$ away from $\lambda=0$ via the unbinding
of, say, dimer A leading to the production of protein B, is not
sufficient to flip the switch: most likely the dimer will rebind the
operator, blocking further production of $B$; only when the dimer A
dissociates {\em and} one of its monomers is degraded will the system commit
itself to the stable state $B$. The probability that the dimer is
degraded before it rebinds the operator increases as the rate of dimer
dissociation increases; 
this is the origin of the increase of $k_{AB}$, $R$ and $\rho(q^*)$
with increasing dimersiation rate $k_{\rm f}$ for low $k_{\rm f}$ seen in
Fig. \ref{fig:sw_rates_GS}. 
Finally, we note that the discrete character of the components in
combination with their low copy number is important
\cite{Shnerb00,Togashi01}: flipping the switch away from the
metastable state at $\lambda=0$
requires the unbinding and subsequent degradation of essentially one
molecule. The metastability is indeed absent in a mean-field continuum
analysis that ignores the discrete nature of the components.


\section{Discussion}\label{sec:dis}

In this paper, we have analysed the stability and switch flipping
dynamics of two types of bistable genetic toggle switches, as a
function of the rates of transcription factor dimerisation and
operator binding. 
 This allows us to assess the
influence of two key sources of fluctuations in the network on the
overall system behaviour.

We have varied the rate constants of the TF-TF and TF-DNA
association/dissociation reactions over more than three orders of
magnitude (see Figs. \ref{fig:sw_rates_ES} and \ref{fig:sw_rates_GS}).
This reflects the wide range of observed rate constants for cellular
biochemical reactions.  For instance, in prokaryotic cells, the
inverse rate of protein production, $k_{\rm prod}^{-1}$, is in the
range seconds to minutes \cite{Record96}. Since the size of a typical
prokaryote is about $1\mu {\rm m}^3$ (based on {\em{E. coli}}), this
corresponds to $k_{\rm prod} V = 10^{-2}-10 {\rm nM}^{-1}/{\rm
  min}$. The rate of monomer-monomer association, $k_{\rm f}$, is
about $10^{-2}-10^{-1} {\rm nM}^{-1}/{\rm min}$, while the dimer
dissociation rate is of the order of $k_d = 10^{-2}-10^{3}/{\rm min}$,
corresponding to dissociation constants in the range $K_{\rm D}^d =
0-10^2 {\rm nM}$ \cite{Buchler05}. This means that $k_{\rm f} =
10^{-2}-10 k_{\rm prod}V$. Figs. \ref{fig:sw_rates_ES}A and
\ref{fig:sw_rates_GS}A show that the switching rate $k_{AB}$ is fairly
insensitive to changes in the dimerisation rate when $k_{\rm f} >
k_{\rm prod}V$, but is highly sensitive to dimerisation rate for
$k_{\rm f} < k_{\rm prod}V$. This shows that the rate of dimerisation
can strongly affect the network stability under biologically relevant
conditions. Rate constants for protein-DNA association/dissociation
are observed to vary over a similarly broad range
\cite{Record96}. Figs.  \ref{fig:sw_rates_ES}B and
\ref{fig:sw_rates_GS}B demonstrate that this variation can have a
marked effect on flipping rates for multistable networks in living
cells.

The steady state phase space density in the region of the stable
states is very robust to every parameter change
(Fig. \ref{fig:Eq}). Yet, changing the rate constants does strongly
affect the switching between these states.  Factorising the switching
rate into the probability $\rho(q^*)$ of finding the system at the
dividing surface, and a kinetic prefactor $R$, we find different
results for the two versions of the switch: while for the exclusive
switch dimerisation affects the switching rate predominantly via the
kinetic prefactor, for the general switch it affects both the kinetic
prefactor and the probability of being at the separatrix; on the other
hand, operator binding affects the flipping rate of the exclusive
switch both via $R$ and $\rho(q^*)$, whereas the effects on the
flipping rate of the general switch are
exerted predominantly through a modification of its steady state
distribution near the separatrix.

These results can be understood by analyzing the transition paths and
the transition state ensemble (TSE). This shows that, in the case of
the exclusive switch, changes in the operator binding rate strongly
affect the properties of the TSE, while the dimerisation rate has
little effect on the TSE.
We conclude that, for the exclusive switch, operator binding
fluctuations play a crucial part in the reaction coordinate, while
dimerisation fluctuations can affect the dynamics of the transition
but have little effect on the route that it takes in phase space. The
case of the general switch is rather different: here, the presence of
the sterile, doubly-bound state $O{\rm A_2B_2}$ makes the flipping
pathways rather insensitive to both sources of fluctuations, even
though the latter do affect the flipping rate. The resolution of this
paradox lies in the fact that the switch is a non-equilibrium system:
in contrast to equilibrium systems that obey detailed balance and
microscopic reversibility, in non-equilibrium systems the forward and
backward transition paths can form a cycle, as observed here; changing
microscopic transition rates (i.e., reaction rate constants) can then change the
stationary distribution $\rho(q)$ and the flipping rate, even though the
location of the transition paths in state space is unaltered.
 Protein-protein and protein-DNA association and
dissociation reactions are a common feature of a wide range of
biological control networks. We therefore hope that our results will
be useful to understand the factors governing stability in multistable
biochemical networks in general.

Genetic switch flipping is an example of a non-equilibrium rare
event. Rather few studies have been made of rare events in
non-equilibrium systems, but a variety of simulation and analytical
approaches have been developed to analyse rare events in equilibrium
systems. Here, it is often assumed that one coordinate, the ``reaction
coordinate'', is slow, while the other degrees of freedom are fast. In
this case, the transition can be modelled by assuming that the
reaction coordinate evolves according to a Langevin equation, while
the other degrees of freedom play the role of friction. Although the
concept of free energy is not applicable to non-equilibrium systems,
one can nevertheless define a ``barrier'' which corresponds to the
maximum of $-\log{\rho(q)}$, as we do in this paper. The results
presented here show that that ``barrier crossing'' in the model toggle
switch differs fundamentally from this classical scenario. For the
genetic switch, the reaction coordinate consists of at least two
parameters, namely the difference in total copy number of species A
and B and the state of the operator \cite{Allen05}. Moreover, these
coordinates evolve on comparable time scales---the operator state
fluctuates on time scales similar to those of protein production and
decay; in addition, their dynamics mix in a non-equilibrium fashion
\cite{Walczak05}---the degradation and production of proteins are
non-equilibrium processes. This hampers the application of standard
theoretical tools to model barrier crossings \cite{Walczak05}. New
theoretical approaches may be required to model such rare events in
non-equilibrium systems.

\begin{acknowledgments}
The authors are grateful to Patrick Warren for many valuable discussions. This work is part of the research program of the
``Stichting voor Fundamenteel Onderzoek der Materie (FOM)", which is financially supported by the "Nederlandse organisatie voor
Wetenschappelijk Onderzoek (NWO)''. R.J.A. is funded by the Royal Society of Edinburgh.
\end{acknowledgments}

\bibliographystyle{biophysj}
\bibliography{sysbio}

\newpage

\begin{center}
{\bf LIST OF FIGURES}
\end{center}

\begin{enumerate}

\item Pictorial representation of the model switch, corresponding to
  reaction scheme (\ref{eq:original_set}). Two divergently-transcribed
  genes are under the control of a shared regulatory binding site on
  the DNA (the operator site $O$). Proteins A and B can bind, in
  homodimer form, to the operator. Each TF acts to block the
  production of the other species. In the exclusive switch, only one
  type of TF can bind at any given time (meaning that never the
  production of both species can be suppressed), whereas in the
  general switch both types of TF can bind (in which case the
  production of both species is repressed).
\label{fig:diagram}

\item
Panels A and B show the switching rate $k_{AB}$ for the exclusive switch as a function of the dimerisation rate $k_{\rm f}$
  (A) and the rate of operator binding $k_{\rm on}$ (B).
  Dissociation rates are scaled such that the equilibrium constants
  remain constant: $K_{\rm D}^d=k_{\rm b}/k_{\rm f}=1/V$ and $K_{\rm
    D}^b=k_{\rm off}/k_{\rm on}=1/(5V)$. Panels C and D show the probability $\rho(q^*)$ of being at the dividing
  surface, as a function of $k_{\rm f}$ (C) and  $k_{\rm on}$ (D). Panels E and F show the kinetic prefactor, as defined by Eq. (\ref{eq:kAB}),
  as a function of $k_{\rm f}$ (E) and  $k_{\rm on}$ (F).
  \label{fig:sw_rates_ES}
  
  \item
Panels A and B show the switching rate $k_{AB}$ for the general switch as a function of the dimerisation rate $k_{\rm f}$
  (A) and the rate of operator binding $k_{\rm on}$ (B).
  Dissociation rates are scaled such that the equilibrium constants
  remain constant: $K_{\rm D}^d=k_{\rm b}/k_{\rm f}=1/V$ and $K_{\rm
    D}^b=k_{\rm off}/k_{\rm on}=1/(5V)$. Panels C and D show the probability $\rho(q^*)$ of being at the dividing
  surface, as a function of $k_{\rm f}$ (C) and  $k_{\rm on}$ (D). Panels E and F show the kinetic prefactor, as defined by Eq. (\ref{eq:kAB}),
  as a function of $k_{\rm f}$ (E) and  $k_{\rm on}$ (F).
  \label{fig:sw_rates_GS}

\item Probability distribution as a function of the order parameter
  $\lambda=n_A - n_B$, with $n_X$ the total copy number of species X,
  for the exclusive switch (A) and for the general switch (B).  The
  distributions are obtained with FFS calculations \cite{Valeriani07},
  for three different sets of parameters.
\label{fig:Eq}

\item Switching paths projected onto the $n_A,n_B$ surface, for three
  different sets of parameters. (A) Paths averaged in the $P_B$
  ensemble for the exclusive switch, where $n_A$ and $n_B$ are
  averaged over configurations with the same value of $P_B$, where
  $n_{\rm X}$ is the total copy number of species X. The forward
  paths, corresponding to transitions from $A$ to $B$ are shown with
  solid lines, while the reverse transitions, from $B$ to $A$ are
  shown with dashed lines.  (B) Switching paths for the general
  switch. In this projection, the paths are highly insensitive to
  variations in parameters. The hypersurface $\lambda\!=\!0$ is
  crossed with a lower total number of $A$ and $B$ molecules.
\label{fig:NANB}

\item A) Exclusive switch: probability that a ${\rm{B_2}}$ dimer is
  bound to the operator, $\langle N_{\rm OB_2} \rangle$, as a function
  of $P_B$ for three different sets of parameters.  The solid
  lines correspond to the transition from $A$ to $B$, while the dashed
  lines corresponds to the reverse transition from $B$ to $A$. B)
  General switch: operator occupancies during the transition from $A$
  to $B$ and vice versa (the empty state $O$ is not shown since it is always
  scarcely occupied), for the {\em baseline} parameter set; the results for
  the other parameter sets are indistinghuisable.
  \label{fig:OB2_lambda} 

\item The probability $p(\lambda)$ for the transition state ensemble
  ($P_B=0.5$) for the transition from $A$ to $B$, where $\lambda = n_A
  - n_B$. Top row A) Exclusive switch; Bottom row B) General switch.
 The probability p($\lambda$) is split into colour-coded
  contributions from the three operator states; the area under each
  histogram gives the probability $\langle N_{\rm OX}\rangle$ that the
  operator is bound to species X (the three areas thus sum to
  unity). The left panels correspond to the system with slow
  dimerisation $k_{\rm f} = 0.1$; the middle panels correspond to the
  system with the base-line parameters; the panels on the right
  corresponds to the system with fast operator binding $k_{\rm on} =
  500$.  \label{fig:histo}

\end{enumerate}

\clearpage

\newpage

\begin{figure}
\begin{center}
\includegraphics[width=14cm]{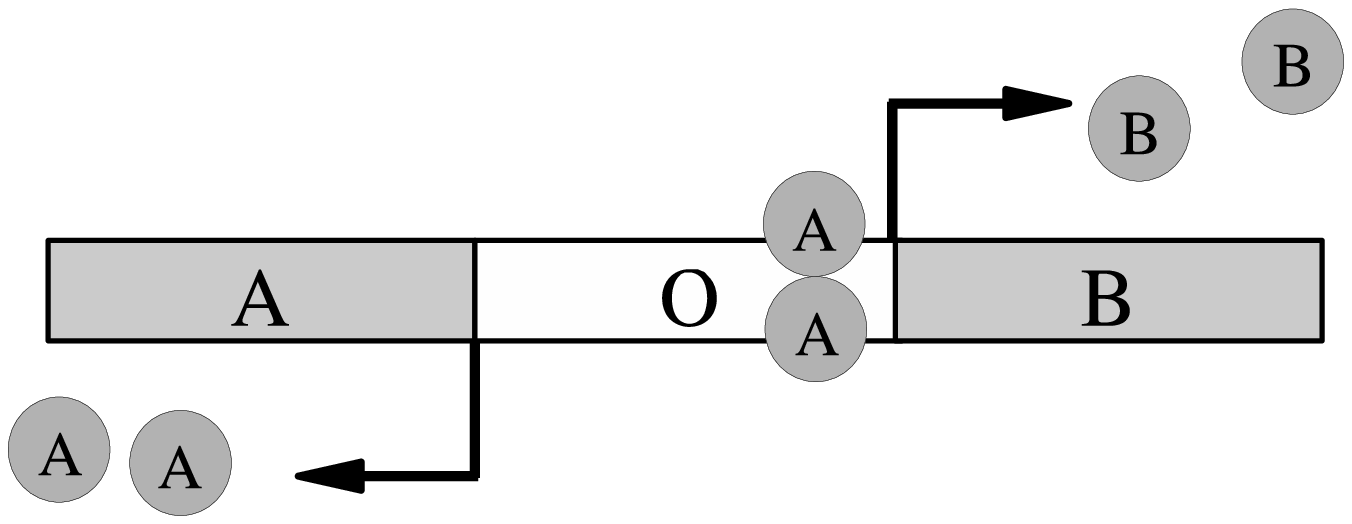}
\caption{Morelli {\em et al.}} 
\end{center}
\end{figure}

\clearpage

\newpage

\begin{figure}
\begin{center}
\includegraphics[width=14cm]{switching_rate_ES.eps}
\caption{Morelli {\em et al.}} 
\end{center}
\end{figure}

\clearpage

\newpage

\begin{figure}
\begin{center}
\includegraphics[width=14cm]{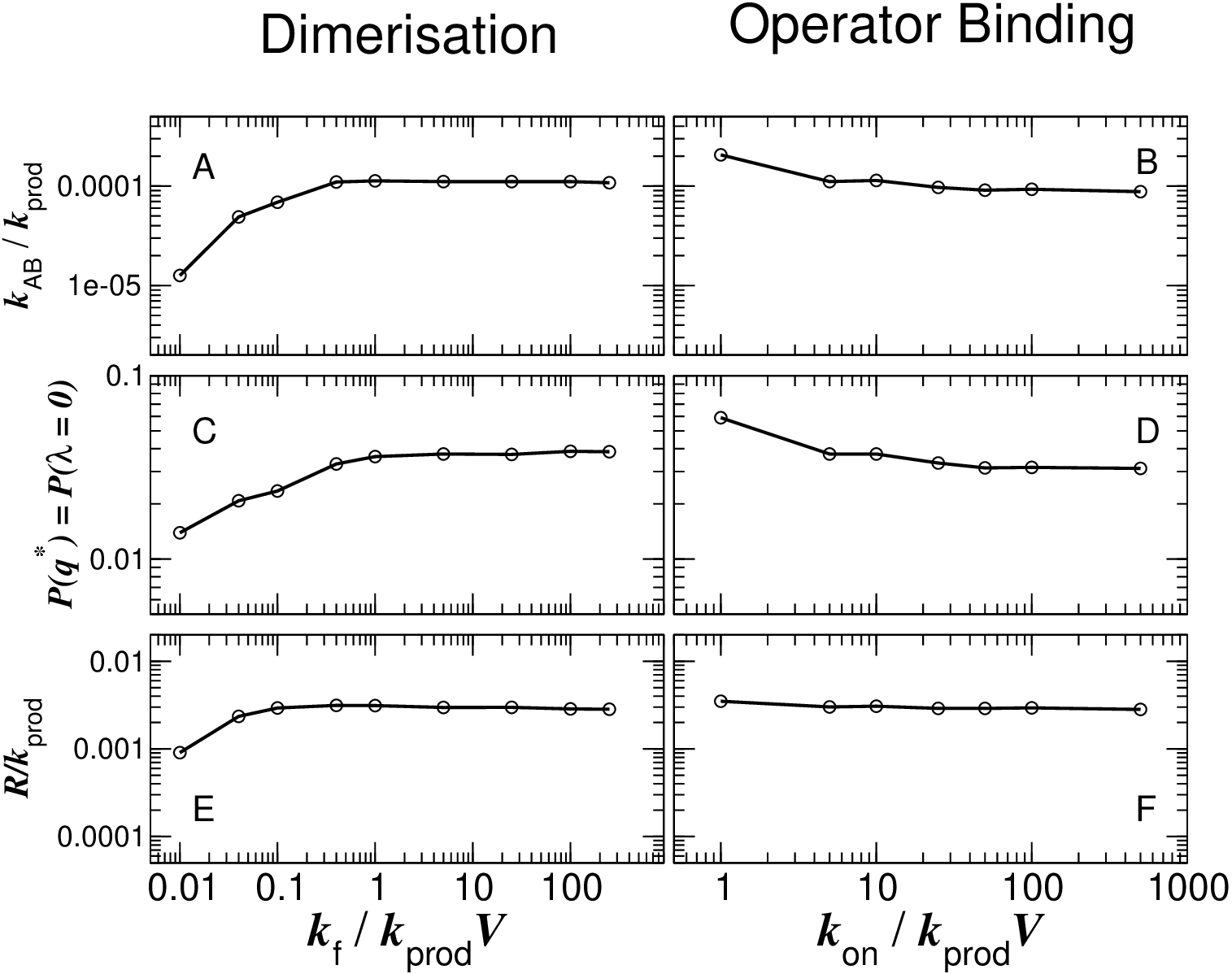}
\caption{Morelli {\em et al.}} 
\end{center}
\end{figure}

\clearpage

\newpage

\begin{figure}
\begin{center}
\includegraphics[width=14cm]{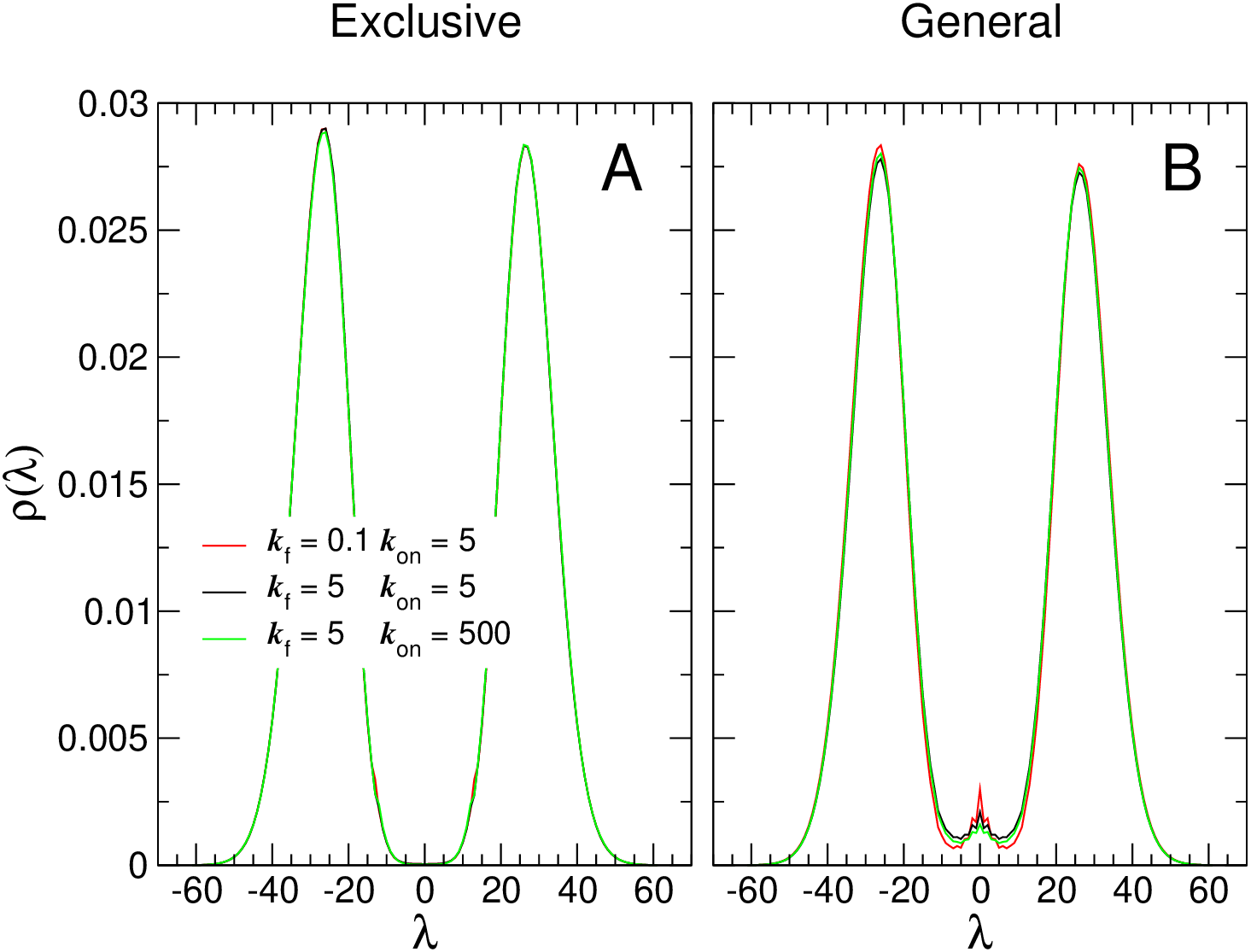}
\caption{Morelli {\em et al.}}
\end{center}
\end{figure}

\clearpage

\newpage

\begin{figure}
\begin{center}
\includegraphics[width=14cm]{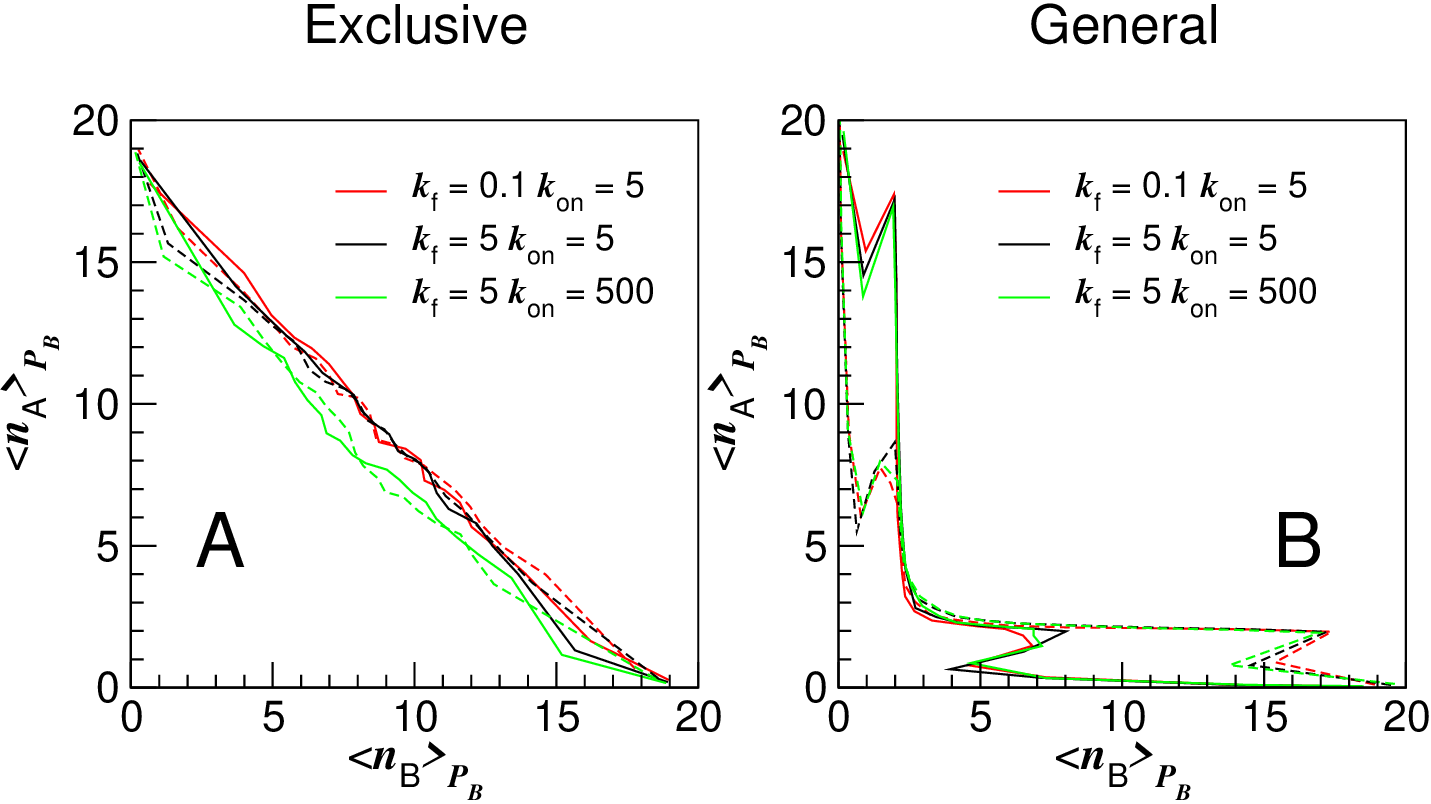}
\caption{Morelli {\em et al.}}
\end{center}
\end{figure}

\clearpage

\newpage

\begin{figure}
\begin{center}
\includegraphics[width=14cm]{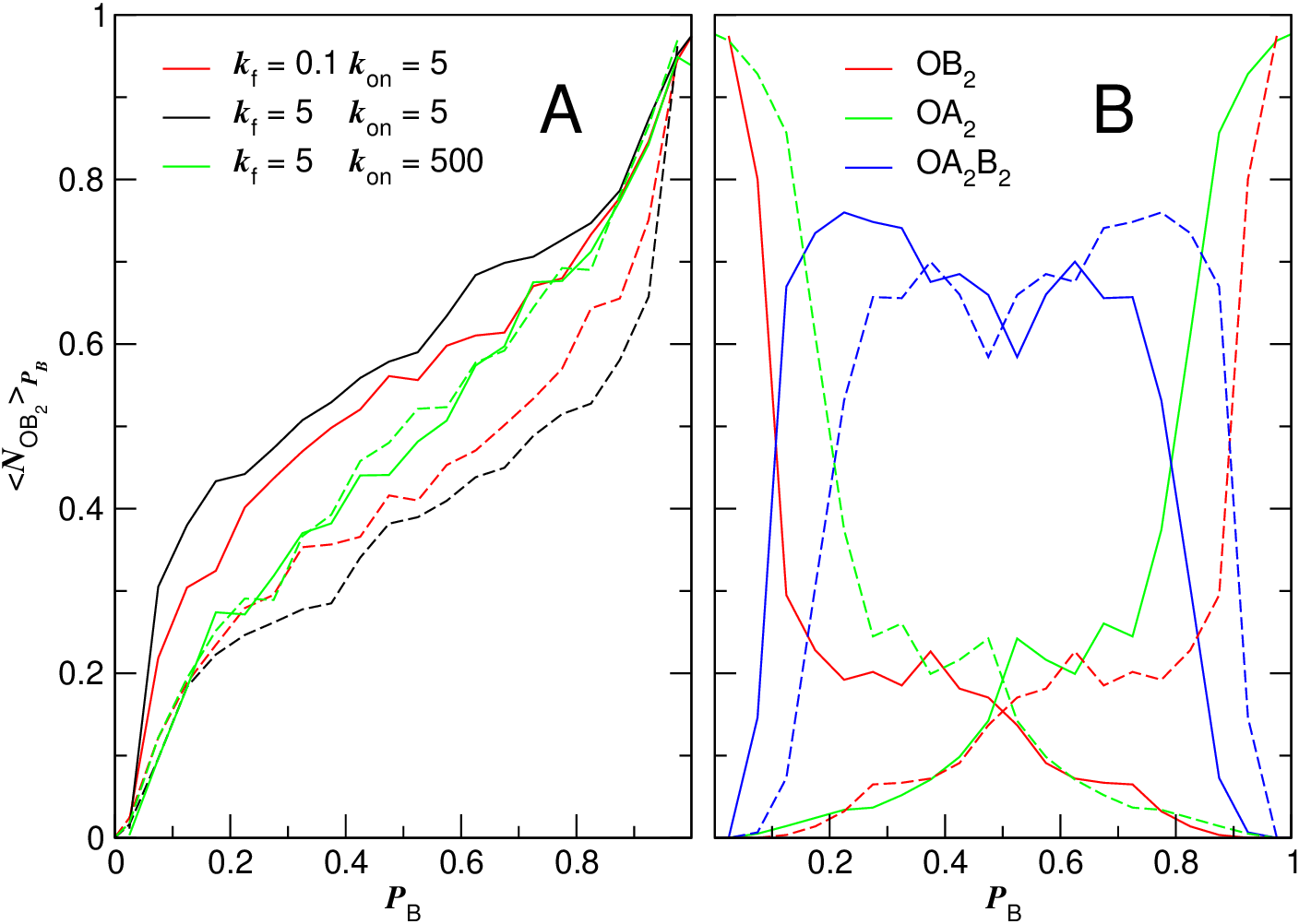}
\caption{Morelli {\em et al.}}
\end{center}
\end{figure}

\clearpage

\newpage

\begin{figure}
\begin{center}
\includegraphics[width=14cm]{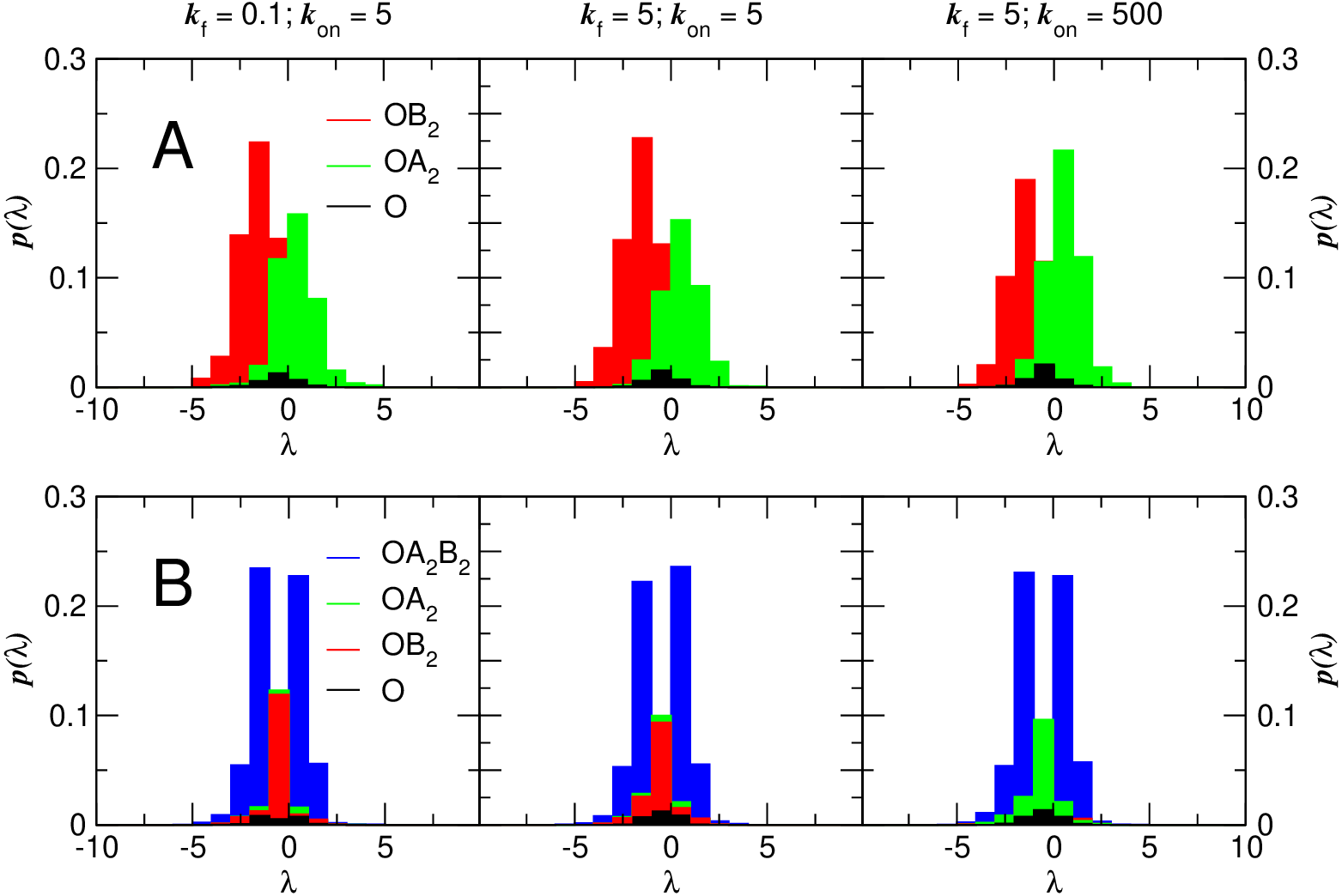}
\caption{Morelli {\em et al.}}
\end{center}
\end{figure}

\clearpage

\end{document}